\documentclass{article}

\usepackage[english]{babel}

\usepackage[letterpaper,top=2cm,bottom=2cm,left=3cm,right=3cm,marginparwidth=1.75cm]{geometry}

\usepackage{amsmath}
\usepackage{graphicx}
\usepackage[colorlinks=true, allcolors=blue]{hyperref}

\usepackage{todonotes}
\setuptodonotes{inline}

\usepackage{acronym} 
\acrodef{SGX}[SGX]{Software Guard Extensions}
\acrodef{TC}[TC]{trusted component}

\usepackage{xcolor}

\xdefinecolor{DarkGreen}{cmyk}{0.8,0,0.8,0.3}

\usepackage{xspace}
\newcommand{\checker}{\textsc{checker}\xspace}
\newcommand{\acc}{\textsc{accumulator}\xspace}

\title{Vivisecting the Dissection:\\ On the Role of Trusted Components in BFT Protocols}

\author{Alysson Bessani$^\ast$, Miguel Correia$^\star$, Tobias Distler$^\dagger$,\\ R{\"u}diger Kapitza$^\dagger$, Paulo Esteves-Verissimo$^\ddagger$, and Jiangshan Yu$^\diamond$\\
\small{$^\ast$LASIGE, Faculdade de Ciências, Universidade de Lisboa, Portugal}\\
\small{$^\star$INESC-ID, Instituto Superior Técnico, Universidade de Lisboa, Portugal}\\
\small{$^\dagger$Friedrich-Alexander-Universit\"at Erlangen-N\"urnberg (FAU), Germany}\\
\small{$^\ddagger$RC3, CEMSE, King Abdullah University of Science and Technology (KAUST), Saudi Arabia}\\
\small{$^\diamond$Monash University, Australia}
}

\date{}

\begin{document}
\maketitle

\begin{abstract}
A recent paper by Gupta et al. (EuroSys'23) challenged the usefulness of trusted component (TC) based Byzantine fault-tolerant (BFT) protocols to lower the replica group size from $3f+1$ to $2f+1$, identifying three limitations of such protocols and proposing that TCs should be used instead to improve the performance of BFT protocols.
Here, we point out flaws in both arguments and advocate that the most worthwhile use of TCs in BFT protocols is indeed to make them as resilient as crash fault-tolerant (CFT) protocols, which can tolerate up to $f$ faulty replicas using $2f+1$ replicas.
\end{abstract}

\section{Introduction}

Byzantine fault-tolerant state machine replication, or just BFT replication, is a method for implementing trustworthy services capable of withstanding arbitrary faults and even intrusions (if combined with other techniques for diversity~\cite{Avizienis85, GarciaB019, Yu23} and resilience~\cite{sousa10prrw}).
These systems have received renewed interest in the last few years due to the emergence of distributed ledger technology.

Much work has been done to design BFT systems capable of achieving high performance and operating with fewer resources.
High performance, in particular with respect to high throughput, is mostly a solved issue, as several works achieve a million or more operations per second in a single small group of replicas (e.g.,~\cite{behl15consensus,behl17hybrids}).
Optimizing resource usage traditionally means decreasing the amount of bandwidth or CPU by devising better algorithms. 
However, in BFT systems, an additional goal is to make algorithms using only $2f+1$ replicas instead of $3f+1$ replicas.
This goal is important because it makes tolerating arbitrarily faulty behavior \emph{as costly as} crash fault-tolerant replication based on widely used algorithms such as Paxos~\cite{Lamport98}, Raft~\cite{OngaroO14}, and Zab~\cite{Jun11}.

Several works target this goal, primarily by using hybrid distributed system models that assume replicas have some \ac{TC} that cannot be corrupted even if the replica is subject to Byzantine failures.
This approach is appealing because relying on a small and arguably simpler \acl{TC}, which should be easier to implement in a trustworthy way than a crash-only full replica, enables the implementation of dependable services with roughly the same amount of resources and performance as crash-tolerant systems.

A recent work by Gupta et al.~\cite{gupta23dissecting} challenges the usefulness of \ac{TC}-based protocols relying on $2f+1$ replicas by (1)~identifying three alleged problems of such protocols (concerning the lack of responsiveness, safety, and parallelism), and (2)~instead proposing trusted components to be primarily used to improve the performance of BFT protocols with \mbox{$3f+1$}~replicas.
From our point of view, both of these contributions contain major flaws that require correction and clarification, which we provide with this paper.
More specifically, concerning~(1), we explain why a closer examination of the existing technical literature would have shown that the responsiveness issue is not specific to the use of trusted components and, furthermore, it can be fixed without sacrificing performance or resorting to a protocol with $3f+1$ replicas.
As for the other two alleged problems, there is strong evidence that they do not exist when the state-of-the-art in this research area is considered.
Concerning~(2), we show that independently of the particularities of any specific implementation, and except for rare corner cases, TC-based protocols with $2f+1$~replicas use available resources more efficiently than TC-based protocols with $3f+1$ replicas and, consequently,  
should provide better performance.

When analyzing the two FlexiTrust protocols that Gupta et al.\ proposed for TC-based BFT systems with $3f+1$~replicas, we identified several parts in which the associated protocol descriptions appear to be underspecified or contain imprecisions.
Among other things, this includes key protocol elements such as the view change and a replica's use of its trusted component, and it reaches an extent to which it is not fully clear whether or not the two protocols are indeed safe.
We document the discovered issues in an effort to advocate for thorough protocol specifications and because they offer an opportunity to study the difficulties of putting the principles of TC-based BFT replication into action.
Ultimately, our aim with this paper is to contribute to a better understanding of the many subtleties involved in designing TC-based protocols.

The remainder of this paper is structured as follows. Section~\ref{sec:hist} provides historical background on architectural hybridization in general and the development of TC-based BFT protocols in particular.
Section~\ref{sec:fundamentals} elaborates on the power of using trusted components to deal with equivocation in hybrid BFT systems, thereby emphasizing the fundamental difference between detecting and preventing equivocation.
Section~\ref{sec:limitations} discusses individually the three alleged problems identified by Gupta et al.\ and revisits them based on the state of the art.
Section~\ref{sec:comparison} analytically compares BFT protocols with $3f+1$~replicas to those with $2f+1$~replicas with regard to their resource usage.
Section~\ref{sec:notes} presents our remarks on the FlexiTrust protocols proposed by Gupta et al., thereby pointing out the importance of precise and comprehensive protocol specifications, especially when it comes to the involvement of trusted components.
Section~\ref{sec:conclusion} concludes the paper and summarizes several additional observations we made during our discussions on this topic.
In Appendix~\ref{sec:tc-designs}, we give an overview of TC implementations to illustrate that the design space in this domain is much larger and more nuanced than it may appear to readers of Gupta et al.'s paper.

\vspace{-1.0mm}

\section{History of TC-based BFT Systems}\label{sec:hist}

\vspace{-1.5mm}

\emph{Trusted component} (TC) is a designation that has been around for years.
A classic example is `trusted third parties' in protocols for secure operations. 
In essence, this would mean that systems with fair behavior concerning security or dependability (untrusted) would somehow feature component(s) whose properties were trusted. 
Trusted components have been used for a long time in modular and distributed systems, with some ambiguity regarding the resulting \emph{model}. The foundation for such mixed behavior~(good and not-so-good components) was not explicitly defined: the system model and architecture remained homogeneous.





With the advent of protocols facing malicious faults -- fault- and intrusion-tolerant (FIT) protocols -- this ambiguity in terms of the model has led to fragilities or vulnerabilities. 
The prototypical example of FIT protocols that inspired most subsequent research are the BFT protocols derived from the seminal work of Castro and Liskov~\cite{castro99practical}.
BFT systems are based on a homogeneous architecture in which computational components are abstracted as processes and networks as channels, supporting a homogeneous system model in the sense that all system components are assumed to have a similar degree of (a)synchrony and be subject to Byzantine failures.

The advent of these “LAN/Internet-facing” protocols, usable in common-day applications, raised enormous interest and led to a research focus on efficiency, complexity, and performance improvements.
The success of these works spawned a significant increment in the use of TC in BFT protocols 
leveraging another event: the progressive availability of ready-made, commercial off-the-shelf (COTS) \emph{hardware-supported trusted components}, such as Trusted Execution Environments (TEEs, e.g., Intel \ac{SGX}, ARM TrustZone, AMD Secure Encrypted Virtualization), Trusted Platform Modules (TPM), and secure coprocessors (IBM Hardware Security Module). 
These components began to appear in several so-called hardware-assisted BFT protocols.
Most of these BFT protocols would be designed, however, in an intuitive way, without necessarily following the guidelines of architectural hybridization theory that are fundamental to the design of correct protocols~\cite{wormholes}.
In fact, be it purpose-made or COTS, the safety properties expected of the TC (i) must be stated, not just implicitly assumed, and (ii) should be part of the assumptions underlying the proof of the payload protocols relying on the TC.
Ignoring this creates a gap that may just be found too late.
For example, PBFT with proactive recovery~\cite{Cas02} was the first BFT protocol to assume a locally trusted component (a synchronous counter).
However, it was discovered later that without a proper architectural setting around such components, it would be impossible to implement resilient proactive recovery~\cite{Sou05}.


\subsection{Architecturally Hybrid Distributed System Models}

The theory of architectural hybridization \cite{TCB2002,wormholes} established the foundation for the correct design of distributed/modular systems based on a few principles, starting with the notion of hybridization.
%
``Good'' components are trusted (e.g., to be secure) because they provide a set of properties (e.g., integrity and confidentiality) despite assumed adversarial behaviors within certain threat models. 
This trust should come from a combination of independence (e.g., they have their own physical or logical capacity, computing, memory, I/O), protection (against tampering and content disclosure), and assurance (verification and validation). 
This intuition is cast in the following principles:



\begin{itemize}
\item[(A)] \emph{Interposition} -- all resources vital to secure the trusted component's properties should be unconditionally controllable by the TC.
No direct access to these resources can be made, bypassing the trusted component(s).

\item[(B)] \emph{Shielding} (tamperproofness) -- the construction of the trusted part is such that it is isolated from the effect of any external fault or attack, including through its interface. 

\item[(C)] \emph{Validation} -- the trusted part should endure a validation process that provides the assurance that it is trusted-trustworthy.
\end{itemize}

In a hybrid model, the different realms follow different fault and/or synchrony assumptions, which require special care in defining the interaction, decoupling, and interface between them:

\begin{itemize}
\item[(D)] The properties of the trusted subsystem (safety, liveness) to be enjoyed by the payload processes should be clearly defined.

\item[(E)] The only way for payload processes to communicate with the trusted component(s) is through a well-defined interface, where the properties of the trusted part are unconditionally obtained, regardless of threats at the untrusted part~\cite{Vuk19}.
\end{itemize}

\subsection{Using Only $2f+1$ Replicas}


BFT algorithms based on hybrid system models can provide several benefits compared to those based on homogeneous system models. An important benefit is reducing the number of replicas needed.
With a homogeneous system model, at least $3f+1$ replicas are needed to mask $f$ being faulty, while with a hybrid system model, only $2f+1$ are needed.
This reduction to $2f+1$ replicas was first achieved with a distributed trusted component called TTCB~\cite{Correia:04a}.
Subsequently, a solution appeared based on a local abstraction called attested append-only memory (A2M) \cite{chun07attested}.
Later, MinBFT, MinZyzyvva~\cite{veronese13efficient} and EBAWA~\cite{Veronese:10} managed to do the same with a much simpler local component called the Unique Sequential Identifier Generator (USIG), similar to TrInc \cite{levin09trinc}.

\subsection{Improving Performance and Scalability}

Previous research has explored various techniques to enhance resilience and performance, such as using A2M, USIG, and TrInc.
These methods employ multiple trusted monotonic counters to deal with equivocation and withstand more faults.
MinBFT~\cite{veronese13efficient} 
and EBAWA \cite{Veronese:10}
provide an improvement over these works (e.g.,~\cite{chun07attested,levin09trinc}) by only using a single trusted monotonic counter to simultaneously improve resilience and reduce the number of phases in PBFT from 3 to 2 during normal case operations.
CheapBFT~\cite{kapitza12cheapbft} further considers the error-free optimistic path where only $f+1$ active replicas are required, and it switches to MinBFT through a fallback mechanism upon detecting faults.
Hybster~\cite{behl17hybrids} proposed TrInX, a trusted monotonic counter similar to TrInc, to enable the parallelization of TEE-empowered BFT instances.
FastBFT~\cite{LiuLKA19} adopted the idea of an error-free optimistic path from CheapBFT while using a TEE-based secret-sharing mechanism for message aggregation instead of threshold-/multi-signature to improve scalability.
Damysus~\cite{DecouchantKRY22} and its chained version are hybrid streamlined protocols with linear communication complexity.
Damysus introduces two trusted components, namely $\checker$ and $\acc$, to independently improve resilience and to
reduce the number of communication phases, respectively.

\section{How Do TC-based BFT SMR Protocols Work?}
\label{sec:fundamentals}

The power of trusted components in BFT systems based on state machine replication~(SMR) is rooted in the fact that they can be used to deal with equivocation, that is, the capability of a faulty process (e.g.,~an elected leader) to make conflicting statements about protocol state~(e.g.,~the request proposed for a certain sequence number or the state with which to initialize a new view) and have these statements accepted by other non-leader processes (followers).
The traditional approach to cope with this problem in BFT~protocols is to introduce an additional protocol phase in which followers exchange information about the leader's proposal and can reveal potential attempts of equivocation.
For example, in PBFT~\cite{castro99practical}, a correct replica only accepts the leader's proposal if at least $2f+1$ (including the leader) of the $3f+1$ replicas back up the same value.
A quorum of such size guarantees that~(within the same view) no other correct follower accepts a different value, meaning that attempts to perform equivocation remain without effect even if they occur.

In general, there are two basic ways to cope with equivocation~\cite{behl17hybrids}: \emph{detecting equivocation} and \emph{preventing equivocation}.
The traditional strategy employed by PBFT is an example of the former category because it allows a faulty leader to potentially misbehave~(by distributing conflicting statements) and relies on followers to later identify and recover from the misbehavior~(e.g.,~by accepting at most one of the conflicting statements and working towards having the leader role assigned to a different replica). 
In contrast, approaches preventing equivocation deprive a faulty leader of the potential to misbehave in the first place and, therefore, require significantly less complex oversight mechanisms for the followers.
Below, we discuss how trusted components have been used to support both approaches.

\paragraph{Detecting equivocation.}
The key to detect equivocation is to assemble some proof that the leader indeed made conflicting statements regarding the same piece of protocol state.
This task involves two parts: (1)~confirming that the statement originates from the leader and (2)~collecting evidence of a conflict. 
Using trusted components, the former is typically done by requiring a leader to authenticate its statements/messages by its local trusted component in the form of \emph{certificates} that only this specific trusted component can produce.
On the other hand, achieving the latter is not as straightforward since a conflict, by definition, involves multiple statements and hence cannot be proven based on a single message alone.
Consequently, a follower must know all the previous leader's statements to detect a conflict.
Protocols such as MinBFT~\cite{veronese13efficient} achieve this by relying on the leader's trusted component to establish an unforgeable and gapless timeline on all statements the leader requests to be certified.
Specifically, for each new certificate, the trusted component creates a unique logical timestamp based on a strictly monotonically increasing counter and includes this timestamp in the certificate.
This way, followers can detect conflicts by obtaining all of the leader's statements since the initial timestamp, verifying the correctness of the corresponding certificates, validating that the sequence of statements contains no gaps, and searching for contradictions in the statements. 
In case a follower finds such a conflict, it has proof that the leader is faulty and ignores all of the leader's further statements beyond this point.

Having made certified but conflicting statements, a faulty leader may try to prevent followers from detecting the equivocation by not disclosing the problematic statements.
Although such behavior can hinder followers from assembling proof of the leader's misbehavior, this does not result in the equivocation attempt being successful.
Instead, due to followers verifying the completeness of the statement sequence provided by the leader, a missing statement will lead to a follower no longer processing any new statements from the leader.
For example, if two conflicting statements~$x$ and~$y$ are certified with timestamps $t_x=47$ and $t_y=55$, followers may advance up to and including $t=46$~(if the leader withholds $x$) or up to and including $t=54$~(if they obtain both $x$ and~$y$, thus detecting the conflict, or if the leader creates a gap by refusing to provide them with $y$). 
In neither case, correct followers accept different statements; hence, the equivocation has no effect.

\paragraph{Preventing equivocation.}
The main idea behind preventing equivocation is to significantly limit the options of a faulty leader by ensuring that it cannot make valid conflicting statements even if it wants to.
To guarantee this, a leader's statements are also certified by its local trusted component to confirm their origin, but instead of a logical timestamp, each certificate includes a unique ID created based on a deterministic and statically defined rule known to all replicas in the system.
Typical inputs for such an ID generation algorithm are the sender's identity as well as the view, sequence number, and protocol phase for which the statement is made~\cite{behl17hybrids} because together, these pieces of information unambiguously define a context that only requires a single statement and never recurs within a system's lifetime. 
Hence, relying on a trusted component to ensure that each ID is attached to at most one local statement effectively prevents the existence of conflicting statements.

Statically defining which ID must be used on which occasion enables followers to check the validity of a statement in isolation, \emph{without requiring any comparison to other statements made by the leader}.
Specifically, a follower is allowed to accept a statement immediately after verifying the corresponding certificate and the correctness of its ID because even if the leader is faulty, there cannot be another certified statement for the same context.

Compared with equivocation detection, the drawback of this approach is the added complexity of the trusted component, which needs to implement the protocol rules mentioned before.
Nonetheless, such added complexity can be very modest and greatly depends on the protocol: Hybster~\cite{behl17hybrids}, for example, uses a significantly simpler TC than Damysus~\cite{DecouchantKRY22}.

\section{Alleged Limitations of Using Trusted Components}
\label{sec:limitations}

In this section, we discuss the three alleged issues that Gupta et al.~\cite{gupta23dissecting} pointed out in existing $2f+1$ TC-based protocols, which they refer to as \textsc{Trust-BFT} protocols, and show that these problems are not inherent to these protocols and can be easily patched.
For each issue, we first present the claim from the introduction of Gupta et al.'s paper, and then elaborate on the potential problem and present a solution without giving up the improved resilience.

\subsection{Responsiveness}

\paragraph{Alleged limitation:} ``\emph{We observe that malicious replicas can
successfully prevent a client from receiving a response for its transactions. 
While the transaction will still commit (consensus liveness), the system will appear to clients as stalled and thus appear non-responsive to clients. 
\textsc{Trust-BFT} protocols allow a reduced quorum size of $f+1$ to commit a request.}''


At the core of the proof presented in support of this claim~\cite{gupta23dissecting}, the client responsiveness cannot be guaranteed as the client needs $f+1$ responses to validate the correctness of the executed operation in BFT SMR.
However, with $2f+1$ TC-based protocols where the quorum is $f+1$, a client can only be guaranteed to receive one response, as the other $f$ replicas that committed the operation can be Byzantine.

\paragraph{Coverage of the problem.} 
The problem affects the protocols of the ``PBFT class" in general, whether they use hybridization or not. 
This limitation is due to the increased relative quorum size required by clients. 
Specifically, it affects PBFT and BFT-SMaRt when using the optimization for executing read transactions without running consensus~\cite{castro99practical} or, more generally, it can affect any protocol in which clients wait for $n-f$ matching replies.


\paragraph{Revisiting the classics.}
The fundamental issue is the mismatch between the requirements for implementing state machine replication and how most BFT protocols actually implement it.
Fred Schneider's seminal tutorial on state machine replication states the following requirement to implement SMR~\cite{Schneider90}: ``\emph{Replica Coordination. All replicas receive and process the same sequence of requests.}''

Intuitively, this requirement translates to clients using a total order multicast~\cite{Had94} to disseminate requests to the replicas.
This primitive satisfies the following Agreement property: \emph{if a correct process delivers $m$, then all correct processes eventually deliver $m$.}
This property is somewhat equivalent to the classical Termination property of consensus: \emph{All correct processes decide some value}~\cite{Had94}.

\paragraph{Why responsiveness is limited.}
Most practical BFT SMR protocols do not implement total order multicast and thus do not satisfy this property.
Instead, systems like PBFT~\cite{castro99practical} and BFT-SMaRt~\cite{BessaniSA14} ensure that (1)~$n-2f = f+1$ correct replicas execute every request and that (2)~eventually all correct replicas have the same state after the system processes a sequence of requests.
Property~1 is ensured by waiting for $n-f = 2f+1$ messages before executing a request, while Property~2 is achieved by using periodic checkpoints and state transfer subprotocols. 

The main consequence of this limitation is that these protocols fail to provide responsiveness when clients need more than $f+1$ matching replies.
This problem was identified in 2021 for PBFT and BFT-SMaRt when using the PBFT's optimization for executing reads without running consensus~\cite{berger21reads}.
This optimization requires clients to wait for $2f+1$ matching replies for normal and read-only operations to preserve linearizability~\cite{castro99practical}.
This also affects other improvements made on SMR protocols, for example, to ensure the database-like durability of operations~\cite{bessani13durable}, where clients need to wait for $n-f$ matching replies (instead of $f+1$) to have the guarantee their operations will be reflected on the system state even if all replicas crash and later recover.

Naturally, this also explains why TC-based protocols using $2f+1$ replicas lose responsiveness: those protocols do not implement the agreement property of total order multicast, and waiting for $f+1=n-f$ matching replies might lead to liveness problems.

Contrary to the statement by Gupta et al.~\cite{gupta23dissecting}, \emph{this is not caused by using weak quorums intersecting in a single process}.

\paragraph{Solutions.}
There are two relatively simple ways to patch existing protocols to solve this implementation problem without increasing the number of required replicas.

\begin{figure}[!t]
    \centering
    \includegraphics[width=1\linewidth]{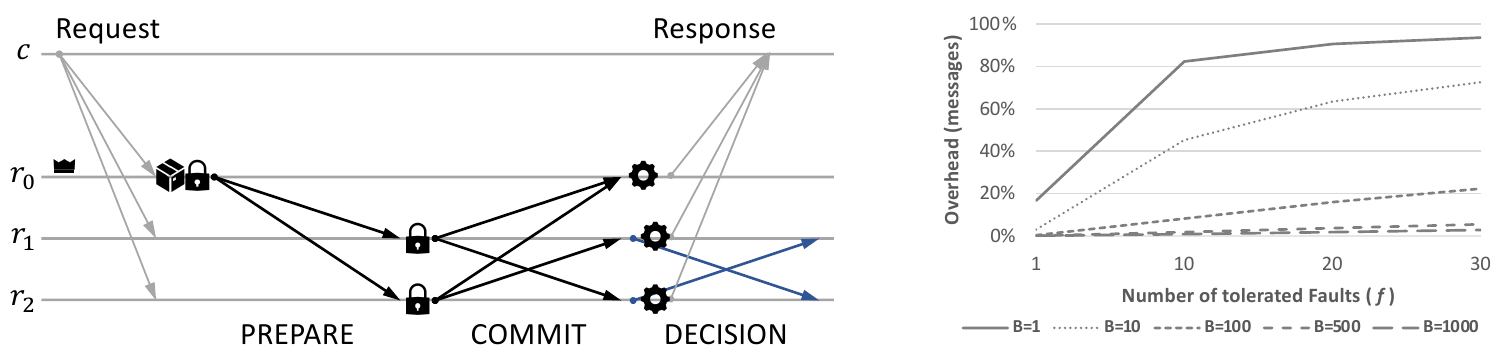}
    \caption{The overhead of Responsive MinBFT. 
    Left: Responsiveness requires \textsc{Decision} messages. 
    Right: The overhead in terms of additional messages per ordered operation for different group sizes and batch values.}
    \label{fig:decision-overhead}
\end{figure}

The first is implementing mechanisms to ensure the Agreement/Termination property.
This can be done using the same design employed in many classical Byzantine consensus protocols (e.g.,~\cite{Baldoni2003,Mostefaoui2015}), in which every deciding process informs the others by broadcasting a \textsc{Decision} message.\footnote{Or, alternatively, making deciding processes participate in at least one additional round of the protocol~\cite{Cachin2001}.}
Gupta et al.~\cite{gupta23dissecting} claim this modification would significantly affect the protocol throughput.
This might or might not be true, depending on how this dissemination is implemented and the batch size used in the system.
However, we argue it is not hard to implement this extension in a very efficient way (see Figure~\ref{fig:decision-overhead}, left) for the following reasons:

\begin{enumerate}

\item Although in the Byzantine setting, such a \textsc{Decision} message would require proof of decision (e.g., $f+1$ TC-signed \textsc{Commit} messages in MinBFT) to trigger a decision in another replica, in theory, such proof can be built using threshold signatures (e.g.,~\cite{Shoup00}), making the size of \textsc{Decision} messages constant.

\item \textsc{Decision} messages do not need to be sent by the leader and neither to the leader.
This is important because it saves the leader's bandwidth, not contributing to the main bottleneck of the protocol, as the leader needs to disseminate message batches.

\item The generation of \textsc{Decision} messages does not need TC access or a signature, and these messages can be completely ignored if the replica has already committed the value.

\item The use of batching -- a requirement for practical BFT protocols and distributed ledgers -- significantly mitigates the overhead of \textsc{Decision} messages.
This happens because the cost of additional messages in the agreement protocol is amortized on the $B$ client-issued operations, with $B$ denoting the batch size.

\item The dissemination of decisions can be delayed for a short time to see if the replica receives some indication other replicas also decided, suppressing the dissemination.
For instance, if a replica sees a \textsc{Commit} for proposal $s+1$ from replica $r$, it knows that this replica already committed proposal $s$ (assuming pipelining is not used).
\end{enumerate}

Figure~\ref{fig:decision-overhead} (right) shows the overhead in terms of additional messages per ordered operation in MinBFT using \textsc{Decision} messages, when compared with the original protocol, for different fault thresholds ($f$) and batch sizes ($B$).
The plots do not consider delayed decisions, as they would trivially make the overhead zero when there are neither faults nor asynchrony (the expected normal case). 
The results show that increasing the number of replicas increases the overhead, as more \textsc{Decision} messages need to be handled, but batching significantly decreases this effect.
For example, with a batch of $500$ messages per agreement, the overhead of the number of additional messages sent per ordered transaction is only $2\%$ with $f=10$ and $5\%$ with $f=30$.
In terms of bandwidth usage (not shown in the figure), with this same batch, the overhead (without considering the threshold signatures optimization described before) is $10\%$ ($f=10$) and $89\%$ ($f=30$) for transactions of $256$ bytes and $3\%$ ($f=10$) and $23\%$ ($f=30$) for transactions of $1k$ bytes.

Another possible solution to the responsiveness issue is to have the leader execute the request before proposing it, and the other replicas verify the outcome~\cite{gueta2019sbft}, as is done in blockchains such as Bitcoin and Ethereum.
In this way, a single reply to the client would contain enough information for the operation to be finalized, namely, the transaction result plus the proof it was approved by $f+1$ other replicas, i.e., the set of $f+1$ TC-signed \textsc{Commit} messages for the request.

\vspace{3mm}
\fbox{\begin{minipage}{14.4cm}
\emph{In summary, although lack of responsiveness is a real issue for $2f+1$ TC-based BFT protocols (and $3f+1$ BFT protocols, in certain cases), the problem is not the use of weak quorums, and it can be easily and efficiently solved by implementing decision forwarding for ensuring termination.}
\end{minipage}}

\vspace{4mm}

\subsection{On Non-trustworthy Trusted Components}
\label{sec:lim-sgx}

\paragraph{Alleged limitation:} ``\emph{Existing \textsc{Trust-BFT} protocols consider an idealised model of trusted computation.
They assume that trusted components cannot be compromised and that their data remains persistent in the presence of a malicious host.
...
A large number of these protocols employ Intel SGX enclaves for trusted computing.
Unfortunately, SGX-based designs have been shown to suffer from rollback attacks...}''


The argument supporting this claim revolves around using SGX technology and its lack of secure, persistent storage. 
In the following, we will delve deeply into why the aforementioned factors fail to offer comprehensive coverage, resulting in assertions that lack complete justification.

In their analysis of recent implementations of trusted subsystems, Gupta et al.\ \cite{gupta23dissecting} discuss how such systems could be susceptible to rollback attacks and, consequently, lose safety.
While early publications utilizing trusted components assumed the support of a trusted platform module where the persistence of counters is secured by the hardware~\cite{veronese13efficient}, subsequent works relying on reconfigurable hardware~\cite{kapitza12cheapbft} or trusted execution environments~\cite{behl17hybrids,DecouchantKRY22} such as SGX left this matter largely unaddressed.
As mentioned in Section~\ref{sec:hist}, the industrial availability of TCs makes them sometimes be used for practical reasons.
However, any effort to put these research works into production requires the assurance that the TC properties are satisfied by the implementation.

\vspace{1mm}

\paragraph{Implementing trusted-trustworthy components.}
The problem of rollback and fork attacks have been prominently discussed and addressed for trusted execution environments such as SGX in ROTE~\cite{rote} and LCM~\cite{LCM}.
Both works state that SGX provides mechanisms to seal and unseal the state of an enclave on disk in an encrypted form to support secure restarts of a system; however, they do not enforce or guarantee freshness.\footnote{It is an example of Shielding violation (see Section~\ref{sec:hist}), as the persistent storage is kept outside the TC.}
In principle, the support for sealing can be combined with trusted counters provided by TPMs, but for frequent state updates, they are too slow, so ROTE and LCM propose witness-based approaches.\footnote{However, recent results significantly improved this situation, e.g.,~\cite{niu22narrator}.}
For TC-based BFT protocols, these solutions would only delegate the problem to a group of external entities and thus are not a self-contained option.

The low performance of TPM trusted counters can be alleviated by request batching as shown in various works (e.g., \cite{veronese13efficient}), but the impact of using a trusted platform model is still severe.
When replicas are deployed in diverse administrative domains, \emph{a requirement for many BFT systems to achieve the necessary level of failure independence}~\cite{GarciaB019, Yu23}, consensus latency significantly increases, diluting the contribution of slower but secure trusted counter access to the end-to-end latency experienced by clients. 
Taking into account the example of TrInc~\cite{levin09trinc} or CheapBFT~\cite{kapitza12cheapbft}, the latter utilizing reconfigurable hardware, it is reasonable to assume that in the process of productization, persistent memory with low access times can be utilized.
Another interesting option, albeit arguably less secure, is to rely on the software-based isolation provided by hypervisors and implement the TC in a small isolated VM tied to each replica~\cite{veronese13efficient}.

\paragraph{Revisiting rollback attacks on Hybster.}
The analysis of Gupta et al.~\cite{gupta23dissecting} focused more on trusted execution environments such as SGX, and here it would be up to major hardware manufacturers to provide the necessary support, which is currently out of reach.
However, the more interesting question might be: is handling persistence as assumed by Gupta et al.\ actually necessary? 
To answer this question, it is worth taking a closer look at TrInX, the trusted component of Hybster~\cite{behl17hybrids}, for which the following assumption is made: ``\emph{As is the case for Intel SGX, we further assume that the execution platform provides means to prevent undetected replay attacks where an adversary saves the (encrypted) state of a trusted subsystem and starts a new instance using the exact same state to reset the subsystem.}'' 
So, how can this be prevented in Hybster and similar systems?
First, the system is set up in a known trusted state during deployment.
This means that the administrator of the Hybster cluster would perform vanilla remote attestation provided by SGX on each TrInX instance and, after establishing a secure connection to the enclave representing the trusted subsystem, inject the shared secret needed to sign messages.
The secret and the counter value will always stay in the TEE and never be persisted on disk.
Thus, classical reboot attacks via a sealed state are not feasible.
Of course, an attacker could stop an enclave and restart it, but according to the system model, this resembles a fault, and the trusted subsystem would not continue to operate.
In summary, the outlined safety issue does not exist; in the worst case, there is a liveness issue if more than $f$ trusted subsystems are stopped in Hybster, which is fine since the system is expected to tolerate at most $f$ faults.
 
However, if we think about productization, what about outages due to maintenance (e.g., a reboot due to applying important security patches)?
In a critical system designed for availability, reboots are commonly a planned procedure.
Thus, as part of the reboot, it can be assumed that the trusted subsystem provides a management interface for the administrator to quiescence it and request a snapshot with the counter.
This can be implemented with standard mechanisms of SGX.
After restarting the system, the administrator performs remote attestation of the fresh, uninitialized TEE and simply reinstalls the saved state, including the shared secret, in a similar way as during initial deployment. 

This leaves the question of unplanned outages, including power failures and system crashes due to software or hardware faults.
If more than $f$ of such failures occur, even if the hardware and main untrusted systems can recover, the trusted subsystems would not, and the replicated service would lose liveness.
However, it should be noted that even non-TC-based BFT systems need to be augmented to provide such type of durability~\cite{bessani13durable}.
In data centers that are designed to host critical services, power outages (hopefully) are prevented by uninterruptible power supplies; however, system crashes would demand root cause analysis, as well as reconfiguring the system to exclude the affected machine and integrate a new one.
Such reconfiguration of a BFT replication group is, for example, supported by BFT-SMaRt~\cite{BessaniSA14} but would require an adaptation for the discussed hybrid systems.

\vspace{3mm}
\fbox{\begin{minipage}{14.4cm}
\emph{In summary, the claim that the missing trusted persistent storage of recent trusted execution environments inevitably leads to safety violations is not justified. As for existing implementations, if more than $f$ trusted subsystems are stopped, liveness could be affected,
which is in line with the expected fault tolerance of the system.}

\hspace{3mm} \emph{
Taking a step back, the problems reported by~\cite{gupta23dissecting} are related to their assumptions how prototypes are implemented, caused by the current use of SGX and limited to this particular technology. In consequence, they are not due to fundamental causes behind the use of TCs in BFT protocols}.
\end{minipage}}

\subsection{Lack of Parallelism}

\paragraph{Alleged limitation:} ``\emph{Existing \textsc{Trust-BFT} protocols are inherently sequential as they require each outgoing message to be ordered and attested by trusted components.}''


This claim posits the impossibility of achieving parallel consensus in TC-based protocols with $2f+1$ replicas.
In the subsequent discussion, we will explore why this assertion is false and elucidate the historical motivations behind the design choices that led to this perception.
%
Furthermore, we discuss three concepts used in traditional non-hybrid BFT protocols to relax the sequential processing of consensus instances, in increasing order of the degree of parallelism that they allow, and explain why these methods can also be applied to hybrid protocols.

\paragraph{Pipelining.}
One approach to speed up the agreement process of a BFT system is to enable subsequent consensus instances to overlap by already starting the first phase of instance~$s+1$ immediately after the end of the first phase of instance~$s$.
Consequently, different phases of different consensus instances typically interleave, enabling a replica to advance one instance while waiting for another to make progress.
For hybrid protocols that follow the principle of \emph{detecting equivocation}~\cite{kapitza12cheapbft,veronese13efficient,distler16resource}, implementing this does not require any special attention because their safety is based on exploiting a gapless sequence of the messages sent by a replica, not on enumerating consensus messages in a certain order~(see Section~\ref{sec:fundamentals}).
In contrast, for protocols \emph{preventing equivocation}~\cite{chun07attested,behl17hybrids}, assigning predefined timestamps to messages is crucial, which is why their designs usually comprise additional measures to enable pipelining.
One effective means to do so is to employ multiple trusted counters, one for each protocol phase.
This way, concerning certifying messages in the trusted component, different protocol phases can be handled independently.
Specifically, a protocol must only ensure that the certificates of each phase are generated in increasing order of consensus instances, which directly fits the underlying idea of pipelining.

The effectiveness of pipelining in exploiting the interleaving of consensus instances depends on the duration of individual protocol phases.
In this context, Gupta et al.\ refer to the round-trip time of each protocol phase as the most important limiting factor~\cite[Sections~1 and 7]{gupta23dissecting}.
This indicates the (implicit) assumption of a HotStuff-like protocol structure such as the one used in Damysus~\cite{DecouchantKRY22}, in which each phase consists of (at least) two communication steps: one for broadcasting a message and the other for collecting the corresponding votes.
As replicas must wait until they have received a sufficient number of votes, this kind of protocol design has a diminishing impact on pipelining effectiveness.
However, HotStuff-style protocols alone do not present the full picture since there are several PBFT-style protocols whose protocol phases comprise only a single communication step.
Among other things, these protocols enable the leader replica to broadcast multiple proposals without any blocking in between and hence decouple the performance achievable by pipelining two consensus instances from the round-trip latency between replicas.
Not surprisingly, this type of architecture serves as the basis for many hybrid protocols~\cite{chun07attested,kapitza12cheapbft,veronese13efficient,distler16resource,behl17hybrids}.

\paragraph{Concurrent consensus instances.}
Although benefiting from pipelining, the traditional PBFT design goes one step further in parallelization.
Specifically, within a limited window of sequence numbers (which is bounded by a low and a high watermark), PBFT allows replicas to run multiple consensus instances in parallel.
Technically, a PBFT leader can even make a proposal for instance~$s+1$ before broadcasting a proposal for instance~$s$. 
Nevertheless, generally, there is no gain in pursuing such a strategy because after the agreement process is complete, client requests still need to be executed in increasing order of their sequence numbers.

Using a PBFT-style protocol architecture, hybrid protocols can exploit the concurrency potential PBFT offers; however, most existing protocols are not designed to do so.\footnote{A notable exception is EBAWA~\cite{Veronese:10}, which not only runs multiple agreement instances in parallel, but does that using different leaders.}
Notice that this is not an inherent limitation of the reliance on trusted components but a consequence of the decision to build trusted components in a specific way.
Namely, to keep their states as small as possible, many trusted components internally only maintain a strictly monotonically increasing counter per timeline to ensure that no logical timestamp (i.e.,~counter value) is assigned more than once.
Relying on a counter for this purpose has the advantage of being a memory-efficient solution, but on the other hand, it introduces an additional restriction.
With the current counter value representing the threshold between used (i.e.,~assigned or skipped) and unused timestamps, new message timestamps must be assigned in increasing order.
For the associated hybrid protocols~\cite{kapitza12cheapbft,veronese13efficient,distler16resource,behl17hybrids}, this creates a dependency between subsequent consensus instances but does not pose a major barrier for parallelization because, as discussed above, it can be effectively addressed by pipelining.
Furthermore, it is worth noting this issue solely concerns the creation of certificates; there are no constraints regarding the order of their verification.

To reiterate, the need to sequentialize the processing of protocol phases results from the goal of having a trusted component design with low complexity, not an intrinsic limitation of hybridization.
At the expense of requiring additional memory, it should be possible to construct trusted components that keep track of the status of individual timestamps within a window, which in turn would offer the opportunity to assign these timestamps in arbitrary order.\footnote{For completeness, it should be mentioned that designing a trusted component this way also requires addressing the problem of allowing a replica to provably skip (a potentially large number of) timestamps, as typically required as part of a view change. For an example, please refer to continuing certificates in Hybster~\cite{behl17hybrids}.}
Providing such a feature is likely to demand additional memory but does not invalidate the approach as a whole.
To our knowledge, such a concept has not been investigated in detail, presumably due to the availability of cleaner and more effective parallelization schemes, as discussed next.

\paragraph{Consensus-oriented parallelization.}
Both approaches discussed so far~(i.e.,~pipelining and concurrent consensus instances) rely, in general, on a single trusted component per replica and, therefore, need to sequentialize the creation of certificates.
If this procedure represents a bottleneck, for example, due to accesses of the trusted component being time-consuming, applying the principle of \emph{consensus-oriented parallelization}~\cite{behl15consensus} makes it possible to introduce full parallelism into a hybrid protocol.
As illustrated by Hybster~\cite{behl17hybrids}, consensus-oriented parallelization separates the agreement process into multiple partitions and afterwards deterministically merges the consensus outcomes of all partitions into a global sequence of committed requests~\cite{distler21byzantine}.
For this purpose, each partition is equipped with its own trusted component~(e.g.,~via instantiation or virtualization), thereby enabling the overall protocol to certify messages of different partitions in parallel.
Experiments with Hybster's trusted subsystem TrInX have shown that for SGX-based components the particular method of instantiation can impact performance~\cite[Figure 5a]{behl17hybrids}.
Specifically, the evaluation indicated a notable decrease in efficiency when multiple instances are provided within the same enclave, and therefore threads from all partitions need to access the same trusted environment concurrently.
In contrast, running each trusted component in its own enclave significantly reduces synchronization overhead and consequently offers good scalability with the number of cores on a server, for example, allowing TrInX to execute more than a million certification operations on just four cores.
Exploiting this potential of distributing load across cores, which is made possible by consensus-oriented parallelization, and combining it with the partition-internal concurrency of consensus instances, hybrid protocols achieve the same degree of parallelism as traditional non-hybrid protocols.

\vspace{4mm}
\fbox{\begin{minipage}{14.4cm}
\emph{In summary, lack of parallelism is not an issue of TC-based BFT protocols. Numerous techniques can accelerate these protocols to the level of the fastest non-hybrid BFT protocols.}
\end{minipage}}

\vspace{2mm}

\section{Comparing $3f+1$ with $2f+1$ Protocols}
\label{sec:comparison}

Hosting replicas is costly for many reasons. 
First and foremost, there is a hardware and infrastructure cost associated with having each replica.
Second, those replicas must be properly managed to guarantee the system's reliability.
Finally, hosting a BFT system requires diversity to ensure that replicas fail independently.
The two later reasons complicate the use of many replicas, as the pool of diversity is typically limited and very costly to maintain~\cite{GarciaB019, Yu23}.
Thus, using fewer replicas to tolerate a certain number of failures is beneficial as it decreases the deployment cost per tolerated failure.

However, Gupta et al.~\cite{gupta23dissecting} pointed out that the cost per transaction of $3f+1$ protocols is better than the cost per transaction of $2f+1$ protocols. 
We believe this is an artifact of their implementation and argue that \emph{the intrinsic characteristics of BFT systems reveal that the resource usage of $2f+1$ protocols will always be lower than $3f+1$ protocols}.
Table~\ref{tab:resources} shows resources used (bandwidth, processing, and TC access) in several leader-based protocols for the expected normal case with no failures and synchrony.
The reported values do not account for client-replica communication and consider the use of signatures instead of authenticators.
This assumption holds also for MinBFT's and Min-ZZ's TC implementations, as the verification of USIG-generated signatures does not require accessing the TC~\cite{veronese13efficient} (see Appendix~\ref{sec:tc-designs}).
Further, we consider an updated version of PBFT that does not require signed \textsc{Commit} messages~\cite{Bravo22}.

\begin{table}[!th]
\centering
\begin{tabular}{|l||c|c|c||c|c|c|}
\hline \hline
Metric          & PBFT           & Flexi-BFT        & MinBFT        & Zyzzyva       & Flexi-ZZ          & MinZZ \\ \hline
Messages handled by leaders & $\approx 6f$   & $\approx 3f$     & $\approx 2f$  & $\approx 3f$  & $\approx 3f$      & $\approx 2f$ \\
Asymmetric crypto/replica$^a$   & $1+2f$         & $1+2f$           & $1+f$         & $1+1$         & $1+1$             & $1+1$ \\
Seq. TC access/consensus  & 0              & $1$ & $2$           & $0$           & $1$  & $2$ \\
\hline \hline
\end{tabular}%
\caption{Usage of resources of several BFT protocols. 
$^a$ Number of signature generations $+$ verifications done by the most busy replicas.}
\label{tab:resources}
\end{table}

The table shows two standard protocols (PBFT and Zyzzyva) and their $2f+1$ (MinBFT and Min-ZZ) and $3f+1$ (Flexi-BFT and Flexi-ZZ, the two protocols proposed by Gupta et al.) counterparts that use TCs.
Let's focus on the MinBFT vs.\ Flexi-BFT case, which captures the most important aspects of the design philosophies under discussion.
Regarding bandwidth usage, the leader replica needs to send $f$ additional proposals, requiring thus $(3f+1)/(2f+1) - 1 = 50\%$ more bandwidth to participate in the consensus protocol in Flexi-BFT than in MinBFT.
The fact that this difference apparently is not reflected in the measurement results in~\cite{gupta23dissecting} indicates that the systems were not evaluated under scenarios that saturate the network; as discussed in Section~\ref{sec:notes}, similar observations can be made with respect to the impact of replies.

In terms of processing, each replica requires two times more cryptographic signature verifications in Flexi-BFT than in MinBFT.
The only metric in which Flexi-BFT is better than MinBFT is the number of TC accesses: in the former, only the leader is required to access the TC, while the latter requires each replica to access the TC once per agreement.
However, these accesses can be fully parallelized using pipelining on the leader proposals (i.e., the leader sends the next proposal as soon as it has enough messages to be ordered, even if the previous proposal is not committed), leading to a modest effect on the overall protocol performance.

Considering the protocols used for the comparison in Gupta et al.~\cite{gupta23dissecting} -- Flexi-ZZ and MinZZ -- the table shows that the only metric with regard to which the former can be better is the number of TC accesses per replica.
However, these accesses are not expected to lead to performance degradation for the aforementioned reasons.

For completeness, it should be noted that under specific circumstances, a larger replica group size can offer some benefits in terms of performance.
In particular, this is the case when costly tasks only need to be performed by a subset of replicas and hence can be load-balanced among the entire group.
One example applies to scenarios where operations' replies are very large, and consequently, the available bandwidth between replicas and clients is a system's limiting factor.
In such cases, the use of hash replies~(i.e.,~only one replica responds with the full result while all others provide result hashes~\cite{castro99practical}) can improve overall throughput if the responsibility for sending the full results is distributed among more replicas.
Another example are approaches that optimistically execute each operation on only a (request-dependent) subset of $f+1$~replicas~\cite{distler11increasing}, which theoretically can improve the request execution throughput by up to $\frac{3f+1}{f+1}$ in systems with $3f+1$ replicas.
As the two examples illustrate, these tasks typically involve client communication or the execution stage and, therefore, are \emph{orthogonal to the use or non-use of trusted components in the agreement stage of a BFT system}.


\section{Notes on the FlexiTrust Protocols}
\label{sec:notes}

Having analyzed the FlexiTrust protocols on a conceptual level in the previous section, in the following, we take a closer look at the specifics of the two protocols proposed by Gupta et al.~\cite{gupta23dissecting}: Flexi-BFT and Flexi-ZZ.
Instead of providing full specifications, both protocols are presented in the form of plain-text discussions and pseudo-code sketches of the protocol phases that are deemed to be most important.
By nature, these kinds of descriptions make it inherently difficult to unequivocally identify potential errors or vulnerabilities, which is why we do not claim to have done so.
Nevertheless, the descriptions contain several imprecisions and omissions that are worth noting, especially because some of them concern the question of how exactly replicas use their trusted components.
Hence, the following analysis also offers insights into the specific problems that need to be addressed when putting the fundamental concepts of TC-based BFT replication into action in the form of concrete protocols.

\paragraph{Counter creation and identities.}

Despite the importance of trusted counters for FlexiTrust protocols, the protocol descriptions of Flexi-BFT and Flexi-ZZ unfortunately provide little information on how counters are created, how the creation of a new counter is announced to other replicas, and how followers verify that the current leader uses the proper counter for its proposals.
The pseudo codes of both protocols assume each replica to have knowledge about the latest counter, however it does not become clear how ``latest'' is defined and how the protocols are able to guarantee that all correct replicas consider the same counter to be active.
On one occasion, the paper mentions that ``\textit{for each new counter that a replica creates, it has to share a certificate (attestation) that proves the newness of this counter}", but simply sharing a certificate does not appear sufficient to ensure that all correct replicas maintain a consistent view on the counter identity.

Notice that a consistent view regarding the identity of the counter currently in use is essential for the FlexiTrust protocols' safety.
Without it, a faulty leader is able to successfully perform equivocation by using different counters for different correct replicas.
This is possible because a trusted component only prevents the same counter value~$k$ of a counter~$q$ from being assigned to more than one message, however, it does not prevent two counters~$q$ and $q'$ from assigning the same counter value~$k$ to different messages.
As a result, scenarios such as the following for example can occur.
A faulty leader creates a new counter~$q$, binds a transaction~$T$ to counter value~$k$, and sends a \textsc{Preprepare} for $T$ to a correct follower.
Immediately afterward, the faulty leader creates a new counter~$q'$, binds a transaction $T'$ to counter value~$k$, and sends a \textsc{Preprepare} for $T'$ to another correct follower.
If both followers consider the respective counters to be valid at the time they receive their \textsc{Preprepare}, they accept the transactions $T$ and $T'$, respectively.
At this point, a fundamental protocol property no longer holds, i.e., the guarantee that within the same view, correct replicas cannot prepare different transactions for the same counter value.
This means that the faulty leader has successfully performed equivocation.

Based on the provided protocol descriptions, it is unclear how the FlexiTrust protocols rule out such a scenario.
As explained in Section~\ref{sec:fundamentals}, in order to be able to prevent equivocation, the recipient of a message needs to have unambiguous prior knowledge based on which it can check the validity of the message.
For the FlexiTrust protocols, this includes an expectation regarding the identity of the used counter, however, the only source for this information appears to be the (potentially faulty) sender of the message~(i.e., the leader).
Protocols such as Hybster~\cite{behl17hybrids} avoid similar problems by using certificate IDs that (among other things) comprise the current view number, which has the benefit of representing a value that is known to all affected replicas without requiring prior actions.
Relying on the trusted components used for the FlexiTrust protocols, such a solution does not seem to be intended, as the presented interfaces do not allow a replica to provide its local trusted component with the necessary information.
Specifically, the method \textsc{Create($k$)} for creating a new counter does not offer the means for a replica to specify the identity~$q$ of the counter, which suggests that the trusted component itself selects~$q$ without any further knowledge of the current protocol state.

\paragraph{View change.}

More than once, Gupta et al.\ put emphasis on the fact that the view-change protocol of Flexi-BFT is ``\textit{identical to the PBFT view-change}''.
On the other hand, the Flexi-BFT's view-change description states that, at the beginning of a view change, replicas include proof about both prepared as well as committed\footnote{Flexi-BFT uses different definitions of the \textit{prepared} and \textit{committed} predicates than PBFT. In Flexi-BFT, a transaction is prepared once a replica accepts the corresponding \textsc{Preprepare} from the leader. For a transaction to be committed, a replica must have obtained $2f+1$~matching \textsc{Prepare}s from distinct replicas; unfortunately, Flexi-BFT's protocol description does not clarify whether or not this quorum is assumed and required to include the \textsc{Preprepare}.} messages in their \textsc{ViewChange} messages to the new leader.
In contrast to PBFT, this means that \textsc{ViewChange} messages may contain \textsc{Preprepare}s that are not backed up by a set of \textsc{Prepare}s, and for which it is thus not clear how the PBFT view-change logic handles them.
Generally, it is not easy to see why the view-change logic designed for a three-phase consensus protocol should be applicable to a two-phase protocol without requiring any modifications.

Apart from the specific handling of messages, from a design perspective, it is also interesting to note that Flexi-BFT apparently does not leverage the new leader's trusted component during view change.
Representing the view-change proposal, a \textsc{NewView} message essentially corresponds to a \textsc{Preprepare} and hence is at risk of being subject to equivocation by a faulty leader.
As a consequence, a view-change design that is tailored to the characteristics of Flexi-BFT might have opened the opportunity to improve view-change robustness.

\vspace{-.5mm}

\paragraph{Reply contents.}

According to the given pseudo code, replicas in both Flexi-BFT and Flexi-ZZ use the same response format to provide their results to clients.
For unknown reasons, the responses include the full original request message~$\langle T \rangle_c$ that a client~$c$ had previously submitted to the service.
The benefits of such a design choice are not obvious and the associate overhead appears to be significant, especially for requests with large transaction payloads.
Consequently, one would expect the FlexiTrust protocols to have a major disadvantage compared with traditional protocols such as PBFT or MinBFT, in which the original request is not part of the reply.
The fact that this effect does not seem to be observable in the experimental evaluation results reported in~\cite{gupta23dissecting} suggests that the workloads used in the experiments were insufficient to present the full picture.

\section{Final Remarks}
\label{sec:conclusion}

This paper re-examined a thought-provoking thesis recently presented by Gupta et al.~\cite{gupta23dissecting}, stating: (i) $2f+1$ trusted component (TC) based BFT protocols allegedly exhibit fundamental limitations affecting their liveness, safety and performance; (ii) ``correct'' alternative designs, of which they give examples in their paper, should use $f$ more replicas, i.e., go back to the classical $3f+1$ replica group size, even using TCs.

\paragraph{Vivisecting the dissection.}
Our in-depth analysis of the reported work showed that, concerning (i), most of the identified issues are, unlike what the authors purport, \emph{implementation} issues, and not \emph{fundamental} ones.
One can be solved with a simple extension of existing protocol implementations, and the other two would not exist in non-naive implementations. 
Concerning (ii), we found problems with FlexiTrust, the protocol proposed as a ``correct'' alternative design using $3f+1$ replicas, further weakening the authors' case. 
Moreover, we explained why, by specification and thus with proper implementation, a state-of-the-art leader-based agreement using $2f+1$ replicas should outperform any similar protocol using $f$ additional replicas.
Further, we point out that prototype implementations and experimental evaluation methodologies for replication protocols can vary greatly, and almost anything can be shown.
So, ``surprising'' conclusions based on experimental results of research prototypes should always be taken with a grain of salt.

\paragraph{The root cause.}
We hope to have constructively shown that care should be taken in mapping and validating concrete implementations of complex distributed protocols versus their root specifications.
This is especially true for TC-based protocols because even when TCs are available on the market (e.g., SGX, TPM), they are `hybrids' under the light of architecturally hybrid distributed systems, whose design principles were briefly recapitulated in Section~\ref{sec:hist}.
Missing those guidelines may bring about problems, as we exemplify below. 

Incidentally, we believe that had FlexiTrust followed the mapping and sanity checks advised by those principles, the problems we noted might have been detected during the design phase.
On the other hand, experimental evaluations of protocol implementations lacking the above contextualization risk arriving at misleading conclusions, taking implementation problems for fundamental issues, as happened in this case. 

Consequently, none of the problems pointed out by the authors afford any conclusion against the effectiveness of properly specified and implemented $2f+1$ hybrid (TC-based) BFT protocols.

\paragraph{Lessons learned.}

On a positive note, the involuntary lesson learned from Gupta et al.~\cite{gupta23dissecting} is that the mismatch between implementations and specifications, undesirable in general, assumes critical proportions in architectural hybridization, in what concerns the hybrid subsystem's properties and the interaction with the payload subsystem, especially when the former are COTS TC. 

When the hybrid is a COTS TC, critical design steps that are often omitted concern the validation of the assumed properties and underlying environment assumptions of the protocol specification part concerning the hybrid versus the reality -- the actual characteristics of the selected hybrid implementation rarely completely match the specification of the particular protocol. 

In fact, the environment may lack adequate support for the required TC, or the TC itself may provide some properties with insufficient coverage (see Section~\ref{sec:hist}). 
For example, we may end up with TCs that are much slower than expected, making the overall performance, or even liveness, of the BFT system much worse than that of a homogeneous system. 
When this is noted at design time, maybe throughput inefficiency can be mitigated, e.g., by amortizing TC access cost with batching. 
However, a latency penalty must be paid (see Veronese et al.~\cite[Section 7.4]{veronese13efficient} for a concrete example). 
Otherwise, that particular TC should not be used for anything other than a proof-of-concept prototype.

Another facet of the problem is ensuring the correctness of the interaction between the payload and the hybrid since these are two different architectural realms.
Using trusted components requires a more complicated setup for the BFT cluster, as secrets and even some small state need to be securely transferred between payload and TC.
This complicates the initialization, recovery, and reconfiguration of the system (see Section~\ref{sec:lim-sgx}) since the properties of hybrids should be unconditionally obtained by payload processes, regardless of threats (see Section~\ref{sec:hist}).
Although some works address these issues for homogeneous BFT systems (e.g.~\cite{bessani13durable,BessaniSA14,Bes20,Duan22}), to the best of our knowledge, they are yet to be addressed in hybrid systems.

In conclusion, we expect our paper to be received as a constructive call to arms for best practices in designing, implementing, and evaluating architecturally hybrid distributed systems and protocols, namely hybrid BFT, when using COTS trusted components.
The lesson is that neglecting this gap leaves mismatches that may defeat the purpose of using hybridization.


\bibliographystyle{alpha}
\bibliography{main}

\appendix

\section{Trusted-Component Designs}
\label{sec:tc-designs}

While trusted components all share the common assumption of only being subject to crash failures, their specific implementations may come in different flavors.
In the following, we present some of the design choices that have been proposed over the years and discuss their properties.

\paragraph{Log vs.\ Counter}

One of the most fundamental design decisions that needs to be made for a trusted component is the abstraction that the component should provide to the (non-trusted) replica.
A2M-PBFT~\cite{chun07attested} for this purpose relies on a log in which the logical timestamps assigned by the trusted component refer to the unique positions of entries in the log.
Once a replica adds a new message to the log, using this abstraction, the trusted component needs to store both (1)~the information that the corresponding entry index is now in use (and hence must not be assigned to another message) and (2)~a hash of the message inserted by the replica.
With message hashes being part of the trusted component state, this approach has the drawback of requiring a significant amount of memory inside the trusted component and means to perform garbage collection by truncating the log.
TrInc~\cite{levin09trinc} addresses this issue by shrinking the abstraction provided by the trusted component down to a counter.
As for the log, the counter values in TrInc are used to assign unique numbers to replica-provided message hashes. 
However, in this case, the hashes are maintained outside of the trusted component, thereby minimizing the size of the component's state.

Notice that the choice of abstraction~(i.e.,~log or counter) is independent of the method applied by a protocol to deal with equivocation~(i.e.,~detection or prevention).
Although existing hybrid protocols commonly employ specific combinations~(e.g.~A2M-PBFT~\cite{chun07attested} prevents equivocation by means of logs, MinBFT~\cite{veronese13efficient} detects equivocation with the help of counters) this does not rule out other variants.
As shown by Levin et al.~\cite{levin09trinc}, it is possible to build a trusted log by managing multiple counters in a trusted component and storing the log contents in non-trusted memory.
However, this type of log implementation (and using completely non-trusted log data structures, for example, to buffer incoming client requests) must not be confused with solutions in which the trusted component provides the entire log abstraction.


\paragraph{HMAC vs.\ Signature}

As discussed in Section~\ref{sec:fundamentals}, the safety of hybrid protocols, to a great extent, relies on the fact that replicas can verify the certificates produced by trusted components.
In particular, this includes (1)~confirming that the certificate indeed stems from the trusted component that is referred to in the certificate and (2)~validating that the certificate is still in its original form and has not been manipulated since.
Typically, this is achieved by implementing certificates based on one of two cryptographic primitives: hash-based message authentication codes~(HMACs) or digital signatures.
Representing a symmetric scheme, the HMAC-based variant uses the same shared key to generate and validate the certificate.
In contrast, the signature-based variant applies an asymmetric approach where certificate creation is done with a secret private key that is only known to the specific trusted component instance, and certificate verification involves a public key that is available to all participating parties.
This difference is crucial as it impacts the interaction between a replica and its local trusted component.
Specifically, HMAC-based certificates require the involvement of the trusted component for both creation and verification, whereas signature-based certificates only need to be generated by the trusted component but can be validated in the non-trusted parts of a replica~\cite{veronese13efficient}.
With regard to performance overhead, this results in a tradeoff that needs to be considered in practice.
For implementations where access to the trusted component is time-consuming, digital signatures can be the more efficient option, even though the necessary computations within the component for digital signatures are usually more expensive than for HMACs.
On the other hand, creating and verifying HMAC-based certificates, in general, are the faster alternative if communication with the trusted component is highly efficient; for example, using CASH~\cite{kapitza12cheapbft}, the durations of such operations are in the range of tens of microseconds.
Consequently, seeking to minimize the number of trusted-component accesses in a protocol~\cite{gupta23dissecting} may, but not necessarily have to, result in the best achievable performance.

\paragraph{Self-selected vs.\ Replica-selected timestamps}

A third design choice is who should select the timestamp to include in a certificate. 
In protocols such as MinBFT~\cite{veronese13efficient} and the FlexiTrust family~\cite{gupta23dissecting}, this task is performed by the trusted component itself by incrementing an internal counter.
That is, apart from (potentially) configuring the initial value of the counter, the non-trusted replica in these protocols does not influence the timestamp used to certify a particular message.
Although not an issue for approaches detecting equivocation (due to the concept requiring a gapless timeline anyway), this property makes it inherently difficult, although not impossible, to support the prevention of equivocation (for which certificates need to comprise predefined timestamps, see Section~\ref{sec:fundamentals}).
As a consequence, protocols like A2M-PBFT~\cite{chun07attested} and Hybster~\cite{behl17hybrids}, for example, opt for trusted-component interfaces that give replicas a certain degree of control over the log-entry indices and counter values that are mapped to messages.
Specifically, this means that replicas can select any timestamp from the pool of timestamps that so far have not been assigned or skipped.
Nevertheless, each trusted component still ensures that no timestamp is used for certifying more than one message, guaranteeing the uniqueness required for safety.

Although suggested otherwise by Gupta et al.~\cite[Section~8.1]{gupta23dissecting}, extending the capabilities of non-trusted replicas in this way for several reasons does not entail additional risks with regard to a Byzantine replica trying to exhaust the timestamp range:

\begin{enumerate}

\item Due to the selection of timestamps solely being relevant for the creation of certificates (not their verification), the possible misbehavior of a faulty replica only has an impact on the replica's local trusted component; that is, the trusted component states of correct replicas in the system remain unaffected by such a procedure.

\item Selecting unreasonably high timestamps is something a Byzantine leader can also do in traditional (non-hybrid) environments, which is why BFT protocols typically restrict the number of concurrent consensus instances to a window limited by low and high watermarks~\cite{castro99practical,distler23micro}.
With hybrid protocols commonly adopting the same concept~\cite{levin09trinc,veronese13efficient,behl17hybrids}, follower replicas simply ignore proposals and other messages if their certificates indicate timestamps far in the future.

\item For protocols relying on equivocation detection, artificially creating gaps in the timeline does not offer benefits for a faulty replica because, as explained in Section~\ref{sec:fundamentals}, such behavior immediately results in other replicas no longer processing messages received from the faulty replica.
\end{enumerate}

Overall, independent of how exactly a Byzantine leader may try to exhaust the timestamp space, if it fails to complete the consensus for new requests, correct replicas will eventually trigger a view change and reassign the role to another replica.
Once demoted to follower, after the view change, the faulty former leader can no longer disrupt the agreement's progress.

\end{document}